# Power-saving Asynchronous Quorum-based Protocols for Maximal Neighbour Discovery

**Mehdi Imani[1,*], Maaruf Ali[2] and Hamid R. Arabnia[3]**

[1]IEEE Member, Stockholm, Sweden
m.imani@gmail.com
[2]Department of Computer Engineering, Epoka University, Vorë, Tirana, Albania
maaruf@ieee.org
[3]Department of Computer Science, University of Georgia, Athens, Georgia, USA
hra@cs.uga.edu
*Correspondence: m.imani@gmail.com

**Abstract:** The discovery of neighbouring active nodes is one of the most challenging problems in asynchronous ad hoc networks. Since time synchronization is extremely costly in these networks, application of asynchronous methods like quorum-based protocols have attracted increased interest for their suitability. This is because Quorum-based protocols can guarantee that two nodes with differing clock times have an intersection within at least one timeslot. A higher neighbour discovery rate of active nodes is desired, but it also results in a higher active ratio and consequently and adversely more overall power consumption of the nodes and a shorter network lifetime. There must be a trade-off between extensive neighbour discovery and active ratio in order to design high-performance and efficient protocols. In this paper, two novel asynchronous quorum-based protocols to maximize the neighbour discovery and minimize the active ratio have been designed and presented. A new metric (Quorum Efficiency Ratio: QER) has also been designed to evaluate and compare the performance of quorum-based protocols in terms of their neighbour discovery (the Expected Quorum Overlap Size: EQOS) and the active ratio. The EQOS has been theoretically derived, along with the Active Ratio and the QER values for the proposed novel protocols and the other contemporary protocols. Finally, the proposed methods have been evaluated and compared against the other methods based on the current metrics and the new metric.

## 1. Introduction

Currently, literally billions of remote and ambient standalone networked sensors are used to monitor our surrounding environment. These sensors can be utilized on an extensive scale to monitor and connect all sorts of devices and transducers in a smart grid system based on an ubiquitous anywhere/anytime/anything style [1-3]. Connecting all these myriad types of devices require an efficient and precise addressing method to effectively address all these objects that belong in the IoT (Internet of Things) universe [4-10]. On the other hand, the generation and conveyance of telecommunication traffic and the growing internet data of the order of 200 exabytes per month [11] are very energy-intensive [12-14]. Moreover, wireless sensor networks (WSNs) normally comprise small sensors with short radio transmission range, limited onboard associated power supply, mediocre processing capability and small data storage resources [15-16]. Limited power supply constitutes the most significant constraint in WSNs [17-18]. Given the practical infeasibility of replenishing energy in situ in the majority of WSN applications, designing energy-efficient systems with longevity is a major challenge.

A commonly used method, quorum sensing, is utilised to design wireless network protocols to prolong the network lifetime. Quorum-based systems mainly rely on sensor nodes for functional switching between units of sensing and communication in an on-off cyclic manner [19-24]. The amount of energy consumption of active nodes can be decreased by more than one-tenth by switching them to sleep mode.

The premise of quorum-based protocols is to divide the time into equal segments known as quorum intervals. Each of these intervals contains $n$ equal beacon intervals during which a station can sleep or stay awake. Quorum systems define cyclic patterns as wakeup-sleep schedules during $n$ consecutive beacon intervals, with the integer $n$ referring to the system size. In addition, the strengths of these protocols lie in the facts that stations ought to be awake in only $O(\sqrt{n})$ out of $n$ beacon intervals with at least two stations remaining awake during every beacon interval. Increasing the number of active slots known as the 'quorum time slots' enhances the likelihood of the forwarder set of nodes being inactive states during their data transmission, which lowers transmission delays. In addition, the number of quorum time slots negatively relates to the lifetime of nodes. Given the challenging design and development of quorum time slots in quorum-based protocols, the available methods currently suffer from limitations such as: poor neighbour discovery; high Active Ratio; working only with fixed array size; using the same quorum size for all the different nodes (non-adaptive methods) and a high end-to-end latency. [25] provides further details and a comprehensive explanation about quorum-based protocols.

In this paper, two new quorum-based protocols are proposed, the 'Adaptive Stepped-Grid' (AS-Grid), for minimising the delay and maximising the neighbour discovery and the 'Low Power Stepped-Grid' (LPS-Grid), for minimising the active ratio and maximising the network lifetime for asynchronous ad hoc networks and wireless sensor networks (WSNs). A new metric, called the 'Quorum Efficiency Ratio' (QER), is defined for evaluating the quorum-based power-saving protocols and all the discussed methods. The AS-Grid was originally presented in [26, 27], but the LPS-Grid and the QER are presented for the first time in this paper. The main novel contributions are as follows:

1) The proposed protocols increase the Expected Quorum Overlap Size (EQOS) [23] and decrease the active ratio to maximize the neighbour discovery and minimise both the delay and power consumption as well. In most existing quorum-based protocols like the grid, torus, e-torus, cyclic and FPP (Finite Projective Plane), the EQOS is low and causes high latency in the sending/receiving of data between the nodes. In this paper, two new quorum systems have been devised that have comparably high EQOS than the grid, cyclic, torus, e-torus and FPP protocols. Another merit of these two protocols is that unlike the existing quorum-based protocols, these two protocols are flexible in handling the system size and works with any array size. While the grid works with just $\sqrt{n} \times \sqrt{n}$ arrays, the torus and the e-torus works with just $t \times w$ arrays when $w = 2t$ and the cyclic and the FPP can only be constructed when $n = k(k-1) + 1$ and $k-1$ is a prime power [23].

2) In the existing quorum systems, the active slots are randomly allocated to nodes in the network. But in the AS-Grid protocol, the active slots are adaptively allocated to the nodes based on the conditions of the nodes. In WSNs, data are collected and transmitted from areas that are far away from the sink or Cluster Heads (CH) nodes to the CH nodes or the sink. Thus, those nodes in areas that are close to the sink or CH nodes need more active slots, while the nodes in areas that are far away from the sink or CH nodes need fewer active slots to transmit the data. This adaptive approach of allocating active slots can consume less energy and prolong the network lifetime [28]. So, an adaptive mode of the AS-Grid is introduced to overcome this problem.

3) The AS-Grid protocol increases the EQOS which thereby decreases the network latency.

4)  The LPS-Grid protocol decreases the active ratio which thereby decreases the power consumption and extends the lifetime of the network.

5)  The results show that the neighbour discovery and power consumption has been improved and furthermore through theoretical analyses also.

Table 1, below, shows a summary of the important acronyms used throughout this paper.

**Table 1.** Summary of Important Acronyms.

| Acronym | Meaning |
|---------|---------|
| ATIM | Ad hoc Traffic Indication Map |
| AS-Grid | Adaptive Stepped-Grid |
| BI | Beacon Interval |
| CH | Cluster Head |
| EQOS | Expected Quorum Overlap Size |
| E-Torus | Extended Torus |
| FPP | Finite Projective Plane |
| IoT | Internet of Things |
| LPS-Grid | Low Power Stepped-Grid |
| MANET | Mobile Ad-Hoc Network |
| MTIM | Multihop Traffic Indication Map |
| PS | Power-Saving |
| QER | Quorum Efficiency Ratio |
| QBP | Quorum-Based Protocol |
| QI | Quorum Interval |
| WSN | Wireless Sensor Network |

The rest of this paper is organized as follows: in Section Two, the problem statement is presented. In Section Three, the related works are reviewed. In Section Four, the details of the AS-Grid and the LPS-Grid protocols are presented. The EQOS, the Active Ratio and the QER are calculated and compared in Section Five. Finally, Section Six presents the conclusions of the paper.

## 2. Problem Statement

Power saving is a crucial problem in Mobile Ad-Hoc Networks (MANET) and WSNs. In these networks, most nodes (sensors) are off-grid, solely relying on battery power, so energy is a scarce resource and must be utilised efficiently. On the other hand, the progress of battery technology has not been fast enough to supply their energy requirements in terms of power density, weight and size. Therefore, intense research work continues to be conducted to propose methods that can make the sensors utilise energy more efficiently. Some of these techniques include power control [29,30]; energy-aware routing protocols [31,32] and power management [33,34].

This research investigates the energy consumption and neighbour discovery problems in an IEEE 802.11-based MANET, which has some special characteristics such as multi-hop communication, mobility and battery dependant power. A power-saving (PS) mode for single-hop MANETs is proposed in IEEE 802.11 [35]. The method is not a perfect fit for multi-hop MANET since the nodes are asynchronous and clock synchronisation is very costly. To overcome this problem, the proposed power-saving protocols for this type of network must be asynchronous. In [36], three asynchronous methods are proposed; among these the quorum-based protocol has gained a lot of attention. This paper identifies some optimal or near-optimal quorum-based systems in terms of the quorum size including the grid [37], the torus [38], the e-torus [39], the cyclic [40] and the Finite Projective Plane (FPP) [37].

Two new quorum techniques are also proposed in this paper: the Adaptive Stepped-Grid (AS-Grid) and the Low Power Stepped-Grid (LPS-Grid). Theoretical analyses are conducted to compare and evaluate the AS-Grid and the LPS-Grid in terms of the active ratio and the EQOS. We also derive an upper bound for the grid quorum system in terms of neighbour discovery. Furthermore, a new metric to evaluate the performance of the quorum systems in terms of the active ratio and neighbour discovery is proposed which also includes carrying out a comparison of all the mentioned methods based on this new metric. In the concept of quorum systems, the lower the active ratio obtained, the higher the performance in terms of the energy efficiency. However, a low active ratio causes a low neighbour discovery rate and consequently leads to a higher end-to-end latency in the network. In order to resolve this problem, a trade-off (or compromise) between the active ratio and neighbour discovery is needed. Therefore, a new metric has been defined, the QER, for evaluating any quorum system based on these limitations.

## 3. Related Works

In quorum-based protocols (QBP), the time is divided into several intervals called the Quorum Interval (QI). Figure 1 shows how each QI is divided into $n$ Beacon Intervals (BI) having the same period. Each BI consists of three windows, the active window, beacon window and the MTIM (Multihop Traffic Indication Map) window. Any BI which starts with a beacon window for that node must compete with the other nodes to send their beacons within this time interval. The beacon packet contains the node address and the node timestamp. After the beacon window, there is an MTIM window that the node waits for receiving the ATIM (Ad hoc Traffic Indication Map) packets of other nodes. The MTIM is analogous to the ATIM in the IEEE 802.11 protocol.

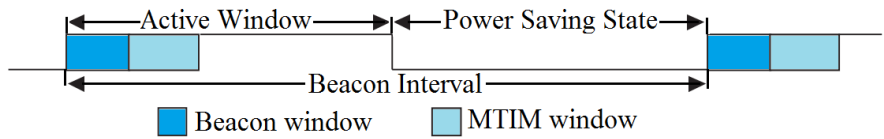

**Figure 1.** The structure of a beacon interval.

In quorum systems, those BIs that belong to 'awake periods' have their associated send/receive windows respectively. Figure 2 shows overlapping slots in a quorum system. At least one overlapping slot between the nodes is guaranteed in this system where the nodes can send/receive their data during the overlapping slots. Each node also independently chooses its BIs and thus no synchronisation bits are needed.

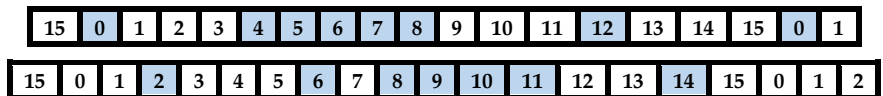

**Figure 2.** Sleep/wakeup periods of a quorum system and the intersection slots (nodes must be awake in the shaded BIs, as shown in the figure).

**Quorum Systems Properties**

**Definition 1**. "Given a universal set $U = \{0. \cdots. n-1\}$, a quorum system Q under U is a collection of non-empty subsets of U, each called a quorum which satisfies the intersection property" [39]:

$$\forall G. H \in Q : G \cap H \neq \emptyset$$

For example Q = {{0,1},{0,2},{1,2}} is a quorum system under U = {0, 1, 2} and each one of these non-empty subsets is a quorum.

**Definition 2**. "Given a non-negative integer $i$ and a quorum H in a quorum system Q under $U = \{0. \cdots. n-1\}$, we define $Rotate(H.i)$" [39]:

$$rotate(H.i) = \{(j+i) \bmod n | j \in H\}$$

**Definition 3**. "A quorum system Q under $U = \{0. \cdots. n-1\}$ is said to have the rotational closure property if" [39]:

$$\forall G. H \in Q. i \in \{0. \dots. n-1\} : G \cap rotate(H.i) \neq \emptyset$$

**Theorem 1**. "If the Q quorum system has rotational closure property, this Q quorum system can be used for solving QPS problem" [39].

**Definition 4**. In a quorum system $Q = \{Q_1. Q_2. Q_3. \cdots. Q_n\}$ under $U = \{0. \cdots. n-1\}$, the active ratio is defined in the relation [39] below:

$$\text{Active Ratio(Qi)} = \frac{|Q_i|}{n} \tag{1}$$

**Theorem 2**. "If Q is a quorum system under $U = \{0. \cdots. n-1\}$ and Q has rotational closure property then each $Q_i$ quorum in Q has at least $\sqrt{n}$ size" [39].

This theorem theoretically defines a lower bound for the size of each quorum with rotational closure property. The proof of this theorem is brought in [39]. In the following, some common quorum systems which satisfy the rotational closure property will be introduced.

### 3.1. Grid Quorum System

Authors in [36] presented a squared quorum system with $\sqrt{n}$ rows and $\sqrt{n}$ columns. In this method, a QI period is mapped to a square with $n$ BIs. Each BI is numbered $(i.j)$, with $i$ being the row number and $j$ the column number. Each host (node) can select a row and a column randomly. The node must be active in all the selected BIs. For example, if $n = 16$, each node contains 16 BIs and each BI is numbered with a pair of $(i.j)$ numbers between 1 and 4. Then, if a node selects row 3 and column 3, then the node must be in these slots: {8,9,10,11,2,6,14}. In this method, at least two overlapped slots are guaranteed between two nodes. Fig. 3 shows two grid quorum systems, G and H, with $n = 16$ that have two intersections in slots 6 and 8. In the grid quorum system, each quorum consists of $(2\sqrt{n}-1)$ slots and consequently, the active ratio is equal to $\frac{2\sqrt{n}-1}{n}$. The grid quorum system is a very simple and commonly used method. But this method just works with squared array sizes that cause its inflexibility in its application.

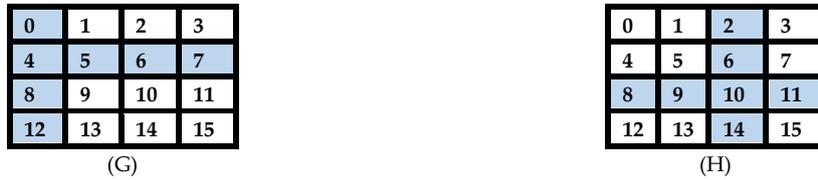

**Figure 3.** Two Grid quorum systems with $n = 16$ that have two intersections in slots 6 and 8

### 3.2. Torus and e-torus quorum systems

In torus [39], like the grid, "the universal set is arranged as a $t \times w$ array where $t \times w = n$. The rightmost column (resp., the bottom row) in the array are regarded as wrapping around back to the leftmost column (resp., the top row). Each node picks up any column c, $0 \leq c \leq w-1$, plus $\lfloor \frac{w}{2} \rfloor$ elements, each of which falls in any position of column $c + i. c = 1 \dots \lfloor \frac{w}{2} \rfloor$" [39]. As shown in [38], if $t = \frac{w}{2}$, the quorum size will be $= \sqrt{2n}$, which is near-optimal. Figure 4 shows a Torus system with $t = 3$ and $w = 6$. As the figure shows, quorum G and H are overlapping in slot 6.

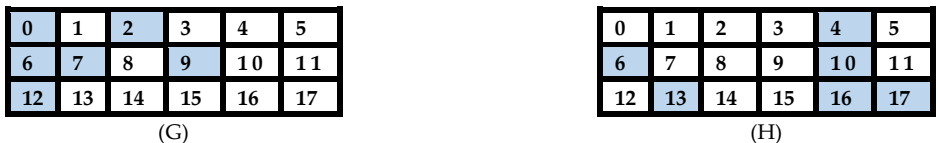

**Figure 4.** Two Torus quorum systems with size $t = 3$ and $w = 6$ that have one intersection in slot 6.

The e-torus (extended torus) quorum system is an extension of the torus quorum system. Like the torus, the universal set $U = \{0. \cdots. t \times w - 1\}$ is arranged as a $t \times w$ array where $t \times w = n$ [23].

**Definition 5**. "Given any integer $k \leq t$, a quorum of an e-torus(k) quorum system is formed by picking any position $[r.c]$, where $0 \leq r < t$ and $0 \leq c < w$, such that the quorum contains all elements on column c plus k half diagonals. These k half diagonals alternate between positive and negative ones and start from the following positions" [23]:

$$\left[r + \left\lfloor i \times \frac{t}{k} \right\rfloor . c \right]. i = 0 \dots k-1$$

"Each quorum in the e-torus(k) quorum system looks like a Christmas tree with a trunk in the middle and k branches, each as a half diagonal, alternating between positive and negative ones" [23]. Figure 5 shows the structure of an e-torus(4) quorum.

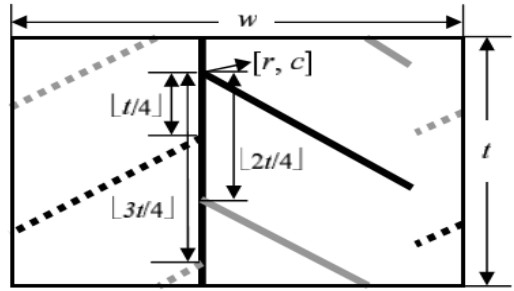

Figure 5. The "Christmas tree" structure of an e-torus(4) quorum [23].

**Definition 6**. "For a quorum system Q under $U = \{0. \cdots . n-1\}$, the Expected Quorum Overlap Size (EQOS) of Q is $\sum_{G,H \in Q} p(G)p(H)|G \cap H|$, where P(G) and P(H) are respectively, the probability of accessing quorums G and H for a quorum access strategy" [23].

### 3.3. Cyclic Quorum System

The cyclic quorum system [40] is formed by the difference sets. A difference set D under $Z_n$ is defined as the following:

$$D = \{d_1. d_2 \ldots d_k\} | \forall e \neq 0 \Rightarrow \exists d_i. d_j \in D \text{ that is } d_i - d_j = e \bmod n. \quad 1 \le i. j \le k$$

To construct a cyclic quorum system with the size of $n$, if D is a difference set under the $Z_n$ universal set, we define Q as below [39]:

$$Q = \{G_1. G_2. \ldots . G_n\}. \text{ where } G_i = \{d_1 + i. d_2 + i. \ldots . d_k + i\}(\bmod n). i = 0. \ldots . n-1$$

"Given any $n$, a difference set as small as $k$ can be found when $k(k-1) + 1 = n$ and $k-1$ is a prime power" [39]. The difference set is called the Singer difference set [41]. For example, the sets {1,2,4} under $Z_7$ and {1,2,4,9,13,19} under $Z_{31}$ are Singer difference sets. Note that in this case, the quorum size $k$ meets the lower bound in Theorem 2. Therefore, a cyclic quorum system that is defined by the Singer difference sets is optimal in terms of the system size [39].

### 3.4. FPP Quorum System

Under the universal set $U = \{0. \ldots . n-1\}$. the FPP [37] is constructed by elements as vertices of a hypergraph. "This hypergraph includes $n$ vertices and $n$ edges. In FPP, each edge is connected to $k$ vertices, and two edges have exactly one common vertex" [37]. A quorum with a size of $k$, can be constructed by the set of all vertices connected by the edge. [37] has shown that an FPP can be constructed when $k(k-1) + 1 = n$ and $k-1$ is a prime power. [40] also shows that the FPP quorum system can be regarded as a special case of the cyclic quorum system when $k(k-1) + 1 = n$ and $k-1$ is a prime power.

### 3.5. Performance Criteria

Two metrics for evaluating quorum systems are shown in this section. A new metric is also defined for evaluating quorum systems in terms of the active ratio and neighbour discovery:

- Active Ratio: the ratio of the active slots in each QI and the length of the QI is called the active ratio [39].
- The Expected Quorum Overlap Size (EQOS): the average number of overlapping slots between two nodes as defined in **Definition 6** precisely. This metric helps us to evaluate the average case neighbour discovery [23].

These metrics are tightly coupled and a slight increase/decrease in one metric can affect the other one. For example, an increasing active ratio causes an increase in the EQOS and vice versa. Traditionally, the active ratio and the EQOS are used to evaluate the performance of the quorum systems. The higher values of EQOS and the lower values of the Active Ratio are desired. However, it is difficult to evaluate the quorum systems by using the active ratio and the EQOS only since these two metrics are the opposite and a proposed method that has a high EQOS but also has high active ratio would not be a practical and efficient method. Some proposed quorum systems have sacrificed the EQOS to get a higher active ratio and vice versa. We need a trade-off between the EQOS and the active ratio. Therefore, we are inspired to define a new metric that includes these two metrics and consider them as a unified metric. Based on this new metric, we can easily evaluate the performance of quorum systems and larger values of this new metric are desirable.

- The QER definition is given below.

**Definition 7.** For a quorum system Q under $U = \{0. \cdots . n - 1\}$, the QER of Q is:

$$QER = \frac{EQOS}{Active\ Ratio} \qquad (2)$$

where the EQOS is defined in **Definition 6** and the Active Ratio is defined in **Definition 4**. This metric considers both the neighbour discovery parameter and the power-saving parameter as a unified metric. The higher the QER, the better the quorum system in terms of its neighbour discovery and power saving.

## 4. Proposed Methods Description

### 4.1. The AS-Grid($t \times w$) Description

We designed a new adaptive quorum system called the Adaptive Stepped Grid (AS-Grid) quorum system, as defined below. The AS-Grid($t \times w$) (for $t \neq w$ or $t = w$) is an adaptive quorum system for use in a wireless sensor network with asynchronous nodes. The AS-Grid is very flexible and easy to implement. The Grid quorum system works with just $\sqrt{n} \times \sqrt{n}$ arrays; Torus works with just $t \times w$ arrays when $w = 2t$, and Cyclic and FPP can only be constructed when $n = k(k - 1) + 1$ and $k - 1$ is a prime power. But the AS-Grid($t \times w$) is very flexible and works with any array size. The AS-Grid satisfies the rotational closure property and is proven in Theorem 3. We derive the Expected Quorum Overlap Size (EQOS) values for the AS-Grid($t \times w$) and compare the results with the EQOS values of all the mentioned quorum systems by analysis. Analytical results show that AS-Grid($t \times w$) has comparably high EQOS values than the FPP, Grid, Cyclic, Torus and E-Torus quorum systems. The AS-Grid is quite efficient and straightforward and can be a suitable alternative for the Grid.

In AS-Grid($t \times w$), a complete QI period is mapped to an array with $n$ beacon intervals, which has $t$ rows and $w$ columns. Each BI is numbered with $(i.j)$, where $i$ is the row number and $j$ is the column number of that BI. But BIs in the AS-Grid($t \times w$) are numbered in a column-major manner, as shown in Figure 6.

**Definition 8.** Given a universal set, in which the elements are arranged as a $t \times w$ array (where $t \times w = n$) with the last column/row in the array regarded as wrapping around back to the first column/row, a quorum of the AS-Grid($t \times w$) quorum system is formed by picking:

(a) all elements of a row $i$, $0 \leq i \leq$ t-1

(b) all elements in column 0, starting from the first BI of QI (slot 0) and ending with the first BI of the selected row.

(c) all elements in column $w - 1$ (last column) starting from the last BI of the selected row and ending with the last BI of QI.

For example, as shown in Figure 6, BIs are selected in the AS-Grid ($3 \times 4$) as the following:

**Figure 6.** Arrangement of all three possible quorum intervals, based on the AS-Grid($3 \times 4$).

For another example to clarify the method, Figure 7 shows the arrangement of all four possible quorum intervals based on the AS-Grid($4 \times 4$). In Figure 7 (a), the selected BIs that a node must be active in them are: {0, 4, 8, 12, 13, 14, 15} and it has intersection with quorum (b) in slots: {0, 13, 14, 15}, with quorum (c) in slots: {0, 14, 15}, and with quorum (d) in slots: {0, 15}.

**Theorem 3**. The AS-Grid quorum system satisfies the rotational closure property.

**Proof.** Let Q be an AS-Grid quorum system. Let H, G ∈ Q. Note that in the AS-Grid($t \times w$) ($w = t \ or \ w \neq t$), there are $t$ different quorum sets. Each quorum set (i.e., H) either has one complete row and one complete column or two incomplete columns that sum up as a full column. Meaning, the number of elements in these two incomplete columns is equal to the number of elements in a full column, covering the width (t) of the AS-Grid (with no gap). Now observe that the rotation of each row in the AS-Grid is another row in the AS-Grid and it follows that $rotate(H.i)$ must intersect with any quorum G ∈ Q because G must contain a full column in the array.

**Figure 7.** Arrangement of all four possible quorum intervals, based on the AS-Grid(4× 4).

In AS-Grid($t \times w$), elements are arranged as a $t \times w$ array, with the last column/row in the array regarded as wrapping around back to the first column/row. In the AS-Grid($t \times w$), all nodes in the same network should choose the same $t$ value. The idea behind that rule is to satisfy the rotational closure property. Using the same $t$ value for all the nodes can guarantee to have an intersection in at least one active slot between two sets with different sizes (see Theorem 4).

In AS-Grid($t \times w$), each node in the same network can choose different $w$ values, but the $t$ values should be the same. Therefore, nodes can change their system size and active ratio based on their remaining energy adaptively. The lower the remaining energy is, the more the system size will be and therefore, the nodes' active ratio will be decreased. For example, node A with a high level of energy can choose AS-Grid(3× 3) and node B with a low level of energy can choose AS-Grid(3× 8). These two quorum systems (with different sizes) have an intersection in at least one active slot. The AS-Grid($t \times w$) is quite simple, easy to implement and efficient, especially in nodes with processing constraints like devices using the Internet of Things.

**Theorem 4**. The AS-Grid($t \times w_1$) quorum system and the AS-Grid($t \times w_2$) quorum system when $w_1 \neq w_2$ satisfies the rotational closure property.

**Proof.** Let H be an AS-Grid($t \times w_1$) quorum system and G be an AS-Grid($t \times w_2$) quorum system when $w_1 \neq w_2$. As H and are using the same $t$, so the proof of this theorem is the same as the proof of Theorem 3.

## 4.2. The LPS-Grid ($t \times w$) Description

We also designed another new adaptive quorum system, called the Low Power Stepped-Grid (LPS-Grid) quorum system as defined below. The LPS-Grid($t \times w$) (for $t \neq w$ or $t = w$) is an adaptive quorum system for use in a wireless sensor network with asynchronous nodes. Like the AS-Grid, the LPS-Grid is also very flexible, easy to implement and works with any array size. The LPS-Grid satisfies the rotational closure property and is proven in Theorem 4. We also derive the EQOS values for the LPS-Grid($t \times w$) and compare our results with the other EQOS values in the theoretical analysis section. The LPS-Grid is quite a simple and energy-efficient solution for asynchronous MANETs or WSNs.

In LPS-Grid($t \times w$), a complete QI period is mapped to an array with $n$ beacon intervals, which have $t$ rows and $w$ columns. Each BI is numbered with $(i. j)$, where $i$ is the row number and $j$ is the column number of that BI. But BIs in the LPS-Grid($t \times w$) are numbered in a row-major manner, as shown in Figure 8.

**Definition 9**. Given a universal set, in which elements are arranged as a $t \times w$ array (where $t \times w = n$) with the last column/row in the array regarded as wrapping around back to the first column/row, a quorum of the LPS-Grid($t \times w$) quorum system is formed by picking:

(a) all elements of a row $i$, $0 \leq i \leq$ t-1

(b) all elements in column $w - 1$ (last column) starting from the last BI of the selected row plus $\left\lfloor \frac{t}{2} \right\rfloor$ BIs of QI (the next sequence of BIs).

For example, as shown in Figure 8, BIs are selected in the LPS-Grid(3 × 4) as follows:

**Figure 8.** Arrangement of all three possible quorum intervals, based on the LPS-Grid(3× 4).

For another example to clarify the method, Figure 9 shows the arrangement of all four possible quorum intervals based on the LPS-Grid(4× 4). In Figure 9 (a), the selected BIs that a node must be active in them are: {0, 4, 8, 12, 13, 14} and it has intersection with quorum (b) in slots: {13, 14}, with quorum (c) in slots: {12, 14} and with quorum (d) in slots: {12, 13}.

**Figure 9.** Arrangement of all four possible quorum intervals, based on the LPS-Grid(4× 4).

In LPS-Grid($t \times w$), like AS-Grid($t \times w$), elements are arranged as a $t \times w$ array, with the last column/row in the array regarded as wrapping around back to the first column/row. The main difference of LPS-Grid($t \times w$) compared to AS-Grid($t \times w$), is that LPS-Grid($t \times w$) has a lower Active Ratio and thereby a lower EQOS than AS-Grid($t \times w$). Like AS-Grid($t \times w$), in LPS-Grid($t \times w$) each node in the same network, can choose different $w$ values, but the same $t$ values to satisfy the rotational closure property.

## 5. Theoretical Analyses

### 5.1. Calculating the EQOS, the Active Ratio and the QER of the Proposed Methods

In this section, we calculate the EQOS, the Active Ratio and the QER of the proposed protocols to evaluate and compare with the other existing protocols. The analysis shows that our proposed methods have better results than the other methods in terms of the EQOS and the QER and also have near-optimal results in terms of the Active Ratio.

### 5.1.1. The EQOS, the Active Ratio and the QER of AS-GRID($t \times w$)

We derive the EQOS of the AS-Grid(5 × 10) and then expand our results to $n$. Figure 10 shows all five possible quorum intervals for the AS-Grid(5 × 10).

**Figure 10.** Arrangement of all five possible quorum intervals, based on the AS-Grid(5× 10).

According to Figure 10, the quorums set of the AS-Grid($5 \times 10$) are as follows:
{a:(0,5,10,15,20,25,30,35,40,45,46,47,48,49), b:(0,1,6,11,16,21,26,31,36,41,46,47,48,49),
c:(0,1,2,7,12,17,22,27,32,37,42,47,48,49), d:(0,1,2,3,8,13,18,23,28,33,38,43,48,49),
e:(0,1,2,3,4,9,14,19,24,29,34,39,44,49)}

If we assume that there are two nodes that want to select one of the quorum sets in a random manner and node 1 selects quorum A, then:

- With the probability of 1/5, the second node selects quorum A and the overlap size, in this case is 14.
- With the probability of 1/5, the second node selects quorum B and the overlap size, in this case, is 5.
- With the probability of 1/5, the second node selects quorum C and the overlap size, in this case, is 4.
- With the probability of 1/5, the second node selects quorum D and the overlap size, in this case, is 3.
- With the probability of 1/5, the second node selects quorum E and the overlap size, in this case, is 2.

Thus, the overlapping average for this case is:

$$A = {^1/_{25}} \ (14+5+4+3+2)$$

If node 1 selects quorum B, as above the overlapping average for this case is $B = {^1/_{25}} \ (14 + 5+5+4+3)$

If node 1 selects quorum C, as above the overlapping average for this case is $C = {^1/_{25}} \ (14+4+5+5+4)$

If node 1 selects quorum D, as above the overlapping average for this case is $D = {^1/_{25}} \ (14+3+4+5+5)$

If node 1 selects quorum E, as above the overlapping average for this case is $E = {^1/_{25}} \ (14+2+3+4+5)$

Summing up the above results for all cases, the EQOS of the AS-Grid($5 \times 10$) quorum system under the universal set {0, ..., 49} is: $\left\lceil {^1/_{25}} \ (28 + 31 + 32 + 31 + 28) \right\rceil = 6$

As seen in Figure 11, the results of the AS-Grid($5 \times 10$) can be divided into four parts with each part of the results formulated and expanded of the above results into $n$ as follows:

**Figure 11.** Analyses of the EQOS in the AS-Grid($5 \times 10$).

Vertical ellipse: $(w + t - 1) \times t$

Horizontal ellipse: $\sum_{i=0}^{t-2}(t - i)$

Big triangle: $\sum_{j=2}^{t} \sum_{i=0}^{t-j}(t - i)$

Small triangle: $\sum_{j=3}^{t} \sum_{i=0}^{j-3}(t - i)$

Summing up the above results for all the cases, the EQOS of the AS-Grid($t \times w$) quorum system under the universal set {0, ..., n−1} is:

$$\frac{t(t + w - 1) + \sum_{j=3}^{t} \sum_{i=0}^{j-3}(t - i) + \sum_{j=2}^{t} \sum_{i=0}^{t-j}(t - i) + \sum_{i=0}^{t-2}(t - i)}{t^2} \qquad (3)$$

And the Active Ratio of the AS-Grid($t \times w$) is:

$$\frac{t + w - 1}{t \times w} \quad if \ t = \frac{w}{2} \quad then \ \frac{3w - 2}{w^2} \qquad (4)$$

Since the AS-Grid($n$) is similar to the AS-Grid($t \times w$) when $t = w = \sqrt{n}$, so the EQOS of the AS-Grid($n$) quorum system under the universal set {0, ..., n−1} is:

$$\frac{\sqrt{n}(2\sqrt{n}-1) + \sum_{j=3}^{\sqrt{n}}\sum_{i=0}^{j-3}(\sqrt{n}-i) + \sum_{j=2}^{\sqrt{n}}\sum_{i=0}^{\sqrt{n}-j}(\sqrt{n}-i) + \sum_{i=0}^{\sqrt{n}-2}(\sqrt{n}-i)}{n} \tag{5}$$

And the Active Ratio of the AS-Grid($n$) is:

$$\frac{2\sqrt{n}-1}{n} \tag{6}$$

The EQOS of the Grid, Torus, Cyclic and FPP are shown in Table 2. As we can see, as the size of $n$ increases, the EQOS of the Grid is getting closer to number 4.

$$\lim_{n\to\infty}\frac{(2\sqrt{n}-1)^2}{n} = \lim_{n\to\infty}\frac{4n-4\sqrt{n}+1}{n} = 4 \tag{7}$$

And the QER of the AS-Grid($t \times w$) is:

$$\frac{w\left[(t^2+tw-t) + \sum_{j=3}^{t}\sum_{i=0}^{j-3}(t-i) + \sum_{j=2}^{t}\sum_{i=0}^{t-j}(t-i) + \sum_{i=0}^{t-2}(t-i)\right]}{t^2+tw-t} \tag{8}$$

### 5.1.2. The EQOS, the Active Ratio and the QER of LPS-GRID($t \times w$)

Similar to the AS-Grid(5× 10), we derive the EQOS of the LPS-Grid(3× 5) and LPS-Grid(4× 6) then expand our results to $n$.

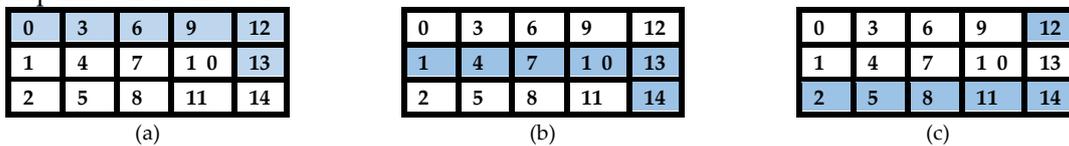

**Figure 12.** Arrangement of all three possible quorum intervals, based on the LPS-Grid(3× 5).

As Figure 12 shows, quorum sets of the LPS-Grid(3× 5) are as follows:

$$\{a:(0,3,6,9,12,13), b:(1,4,7,10,13,14), c:(2,5,8,11,14,12)\}$$

If we assume that there are two nodes that want to select one of the quorum sets randomly and node 1 selects quorum A, then:

- With the probability of $^1/_3$, the second node selects quorum A and the overlap size, in this case, is 6.

- With the probability of $^1/_3$, the second node selects quorum B and the overlap size, in this case, is 1.

- With the probability of $^1/_3$, the second node selects quorum C and the overlap size, in this case, is 1.

Thus, the overlapping average for this case is:

$$A = \,^1/_9 \,(6+1+1)$$

If node 1 selects quorum B or quorum C, then the overlapping average for these cases is the same:

$$B \; or \; C = \,^1/_9 \,(6+1+1)$$

Summing up the above results for all the cases, the EQOS of the LPS-Grid(3× 5) quorum system under the universal set {0, ..., 14} is: $\left[^1/_9\,(8+8+8)\right] = 2.66$

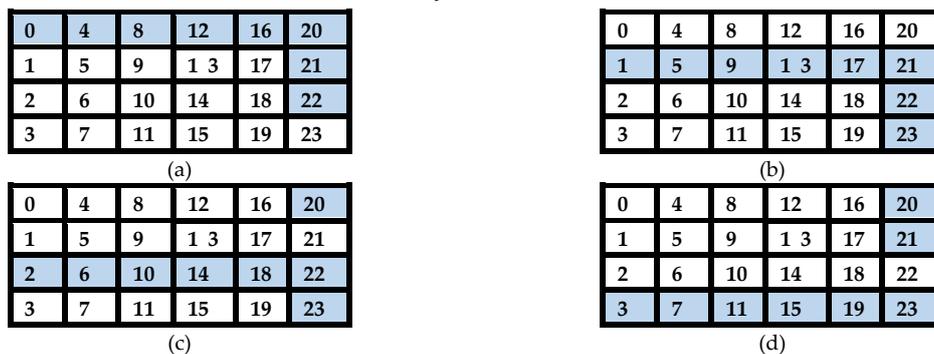

**Figure 13.** Arrangement of all four possible quorum intervals, based on the LPS-Grid(4× 6).

For another example, as Figure 13 shows, quorums set of the LPS-Grid(4× 6) are as follows: {a:(0,4,8,12,16,20,21,22), b:(1,5,9,13,17,21,22,23), c:(2,6,10,14,18,22,23,20), d:(3,7,11,15,19,23,20,21)}

Similarly, if we assume that there are two nodes that want to select one of the quorum sets in a random manner and node 1 selects quorum A, then:

- With the probability of $^1/_4$, the second node selects quorum A and the overlap size in this case is 8.
- With the probability of $^1/_4$, the second node selects quorum B and the overlap size, in this case, is 2.
- With the probability of $^1/_4$, the second node selects quorum C and the overlap size, in this case, is 2.
- With the probability of $^1/_4$, the second node selects quorum D and the overlap size, in this case, is 2.

Thus the overlapping average for this case is:

$$A = \,^1/_{16}\,(8+2+2+2)$$

If node 1 selects quorum B, C, or D, the overlapping average for these cases are the same:

$$B \; or \; C \; or \; D = \,^1/_{16}\,(8 + 2+2+2)$$

Summing up the above results for all the cases, the EQOS of the LPS-Grid(4× 6) quorum system under the universal set {0, …, 23} is: $\left[^1/_{16}(14 + 14 + 14 + 14)\right] = 3.5$

As can be seen in Figure 14, the results of the AS-Grid(5 × 10) may be divided into four parts and the results can be formulated and expanded using the above results into $n$ as follows:

**Figure 14.** Analyses of the EQOS in the LPS-Grid(4× 6)

The vertical ellipse: $\sum_{i=1}^{t}(w + \lfloor\frac{t}{2}\rfloor)$

The big ellipse: $2\sum_{i=1}^{t-1}(\lfloor\frac{t}{2}\rfloor)$

The small ellipses: $2\sum_{i=1}^{t-2}\lfloor\frac{t}{2}\rfloor$

The small squares: $2\sum_{i=1}^{t-3}\lfloor\frac{t}{2}\rfloor$

Summing up the above results for all the cases, the EQOS of the LPS-Grid($t \times w$) quorum system under the universal set {0, …, n–1} is:

$$\frac{\sum_{i=1}^{t}\left(w + \lfloor\frac{t}{2}\rfloor\right) + \, 2\sum_{i=1}^{t-1}\left(\lfloor\frac{t}{2}\rfloor\right) + 2\sum_{i=1}^{t-2}\lfloor\frac{t}{2}\rfloor + 2\sum_{i=1}^{t-3}\lfloor\frac{t}{2}\rfloor}{t^2} \qquad (9)$$

$$where \; t = 3 \; or \; t = 4$$

After simplifying Relation 9, it summarises to Relation 10:

$$\frac{t \times w + 7t\lfloor\frac{t}{2}\rfloor - 12\lfloor\frac{t}{2}\rfloor}{t^2} \; where \; t = 3 \; or \; t = 4 \qquad (10)$$

And the Active Ratio of the LPS-Grid($t \times w$) is:

$$\frac{w + \lfloor\frac{t}{2}\rfloor}{t \times w} \qquad (11)$$

And the QER of the LPS-Grid($t \times w$) is:

$$\frac{w[tw + 7t\lfloor\frac{t}{2}\rfloor - 12\lfloor\frac{t}{2}\rfloor]}{t(w + \lfloor\frac{t}{2}\rfloor)} \; where \; t = 3 \; or \; t = 4 \qquad (12)$$

## 5.2. Performance Evaluation

In this section, the performance evaluation is presented by comparing the performance of the two proposed protocols with the other quorum-based protocols in terms of the EQOS, the Active Ratio and the QER (the proposed metric). The EQOS, the Active Ratio and the QER of the AS-Grid($t \times w$) and LPS-Grid ($t \times w$) were derived and shown in the previous section. Table 2 shows the EQOS, the Active Ratio and the QER values for the quorum-based protocols that were discussed above.

**Table 2.** The EQOS, the Active Ratio and the QER values for the quorum-based protocols.

| | | |
|---|---|---|
| **Grid** | EQOS [23] | $$\dfrac{\left(2\sqrt{n}-1\right)^2}{n}$$ |
| | Active Ratio [23] | $$\dfrac{2\sqrt{n}-1}{n}$$ |
| | QER | $2\sqrt{n}-1$ |
| **Torus** | EQOS [23] | $$\dfrac{\left(t+\frac{\left\lfloor\frac{w}{2}\right\rfloor}{t}\right)+2\left(\left\lfloor\frac{w}{2}\right\rfloor-1\right)\left(1+\frac{\left\lfloor\frac{w}{2}\right\rfloor}{2t}\right)+2}{w}=2$$ |
| | Active Ratio [23] | $$\dfrac{\sqrt{2tw}}{t\times w}\quad if\ t=\frac{w}{2}\ then\ \frac{\sqrt{2}}{\sqrt{n}}$$ |
| | QER | $\sqrt{2tw}$ |
| **Cyclic** | EQOS [23] | $$\dfrac{sn+\lambda\binom{n}{2}}{\binom{n+1}{2}}=\dfrac{2s+\lambda(n-1)}{n+1}$$ where s is the quorum size. |
| | Active Ratio [23] | $$\dfrac{1}{\sqrt{n}}$$ when $k(k-1)+1=n$ and $k-1$ is a prime power. |
| | QER | $$\dfrac{\sqrt{n}[2s+\lambda(n-1)]}{n+1}$$ where $k(k-1)+1=n$, $k-1$ is a prime power and s is the quorum size. |
| **FPP** | EQOS [23] | $$\dfrac{sn+\binom{n}{2}}{\binom{n+1}{2}}=\dfrac{2s+n-1}{n+1}$$ where s is the quorum size, s(s−1) + 1 = n and s−1 is a prime power. |
| | Active Ratio [23] | $$\dfrac{1}{\sqrt{n}}$$ It is shown in [40] that the FPP quorum system can be regarded as a special case of the cyclic quorum system when $k(k-1)+1=n$ and $k-1$ is a prime power. |
| | QER | $$\dfrac{\sqrt{n}[2s+n-1]}{n+1}$$ where s is the quorum size, s(s−1) + 1 = n and s−1 is a prime power. |
| **AS-Grid** $(t \times w)$ | EQOS | $$\dfrac{t(t+w-1)+\sum_{j=3}^{t}\sum_{i=0}^{j-3}(t-i)+\sum_{j=2}^{t}\sum_{i=0}^{t-j}(t-i)+\sum_{i=0}^{t-2}(t-i)}{t^2}$$ |
| | Active Ratio | $$\dfrac{t+w-1}{tw}\ if\ t=\frac{w}{2}\ then\ \dfrac{3w-2}{w^2}$$ |
| | QER | $$\dfrac{w[(t^2+tw-t)+\sum_{j=3}^{t}\sum_{i=0}^{j-3}(t-i)+\sum_{j=2}^{t}\sum_{i=0}^{t-j}(t-i)+\sum_{i=0}^{t-2}(t-i)]}{t^2+tw-t}$$ |
| **AS-Grid** $(\sqrt{n} \times \sqrt{n})$ | EQOS | $$\dfrac{\sqrt{n}(2\sqrt{n}-1)+\sum_{j=3}^{\sqrt{n}}\sum_{i=0}^{j-3}(\sqrt{n}-i)+\sum_{j=2}^{\sqrt{n}}\sum_{i=0}^{\sqrt{n}-j}(\sqrt{n}-i)+\sum_{i=0}^{\sqrt{n}-2}(\sqrt{n})}{n}$$ |
| | Active Ratio | $$\dfrac{2\sqrt{n}-1}{n}$$ |
| | QER | $$\dfrac{\sqrt{n}(2\sqrt{n}-1)+\sum_{j=3}^{\sqrt{n}}\sum_{i=0}^{j-3}(\sqrt{n}-i)+\sum_{j=2}^{\sqrt{n}}\sum_{i=0}^{\sqrt{n}-j}(\sqrt{n}-i)+\sum_{i=0}^{\sqrt{n}-2}(\sqrt{n})}{2\sqrt{n}-1}$$ |
| **LPS-Grid** $(t \times w)$ | EQOS | $$\dfrac{tw+7t\left\lfloor\frac{t}{2}\right\rfloor-12\left\lfloor\frac{t}{2}\right\rfloor}{t^2}$$ where t = 3 or t = 4 |
| | Active Ratio | $$\dfrac{w+\left\lfloor\frac{t}{2}\right\rfloor}{tw}$$ |
| | QER | $$\dfrac{w[tw+7t\left\lfloor\frac{t}{2}\right\rfloor-12\left\lfloor\frac{t}{2}\right\rfloor]}{t(w+\left\lfloor\frac{t}{2}\right\rfloor)}\ where\ t=3\ or\ t=4$$ |

### 5.2.1. The EQOS

The EQOS is used as an average-case value of neighbour discovery. Using the average-case of neighbour discovery to evaluate different quorum systems is more reasonable and accurate than using the worst-case and the best-case values. Figure 15 shows the comparison of the discussed quorum protocols in terms of the EQOS. The results show that the AS-Grid($3 \times w$) and LPS-Grid($3 \times w$) (our proposed methods) have better EQOS than other quorum-based protocols and consequently have better neighbour discovery than the other methods.

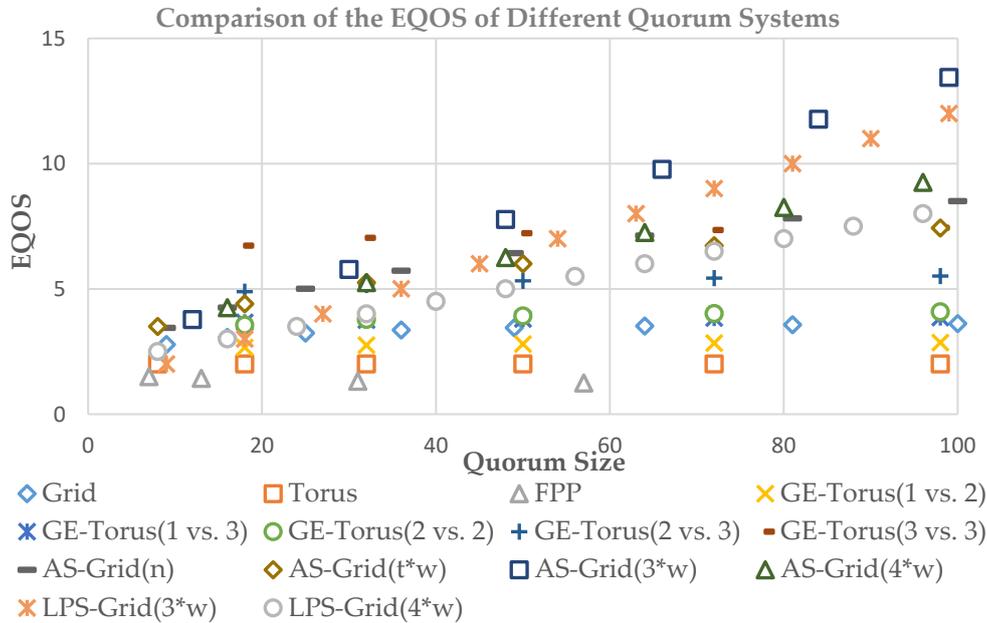

**Figure 15.** The comparison of the EQOS of the discussed quorum protocols.

### 5.2.2. The Active Ratio

The FPP is used as a benchmark. The FPP quorum system, when available, provides the optimal solution [39]. As Figure 16 shows, the LPS-Grid has a near-optimal Active Ratio and the Active Ratio of the AS-Grid is exactly the same as the Active Ratio of the Grid.

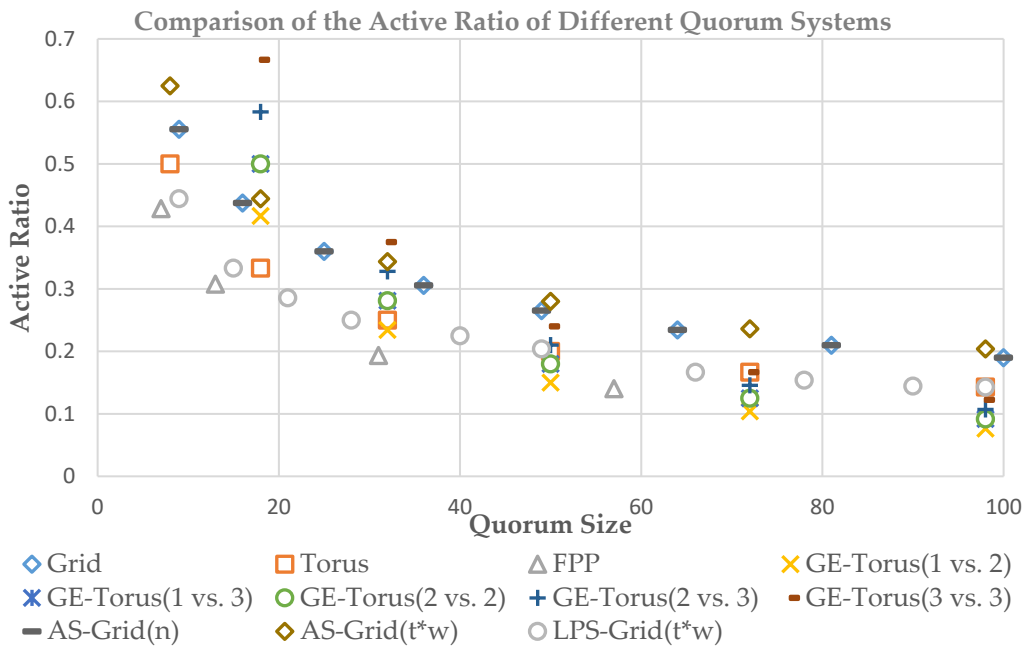

**Figure 16.** The comparison of the Active Ratio of the discussed quorum protocols.

The Active Ratio is a metric to evaluate the methods in terms of power consumption, but it has not provided any information about the neighbour discovery. One method can have an optimal or at least near-optimal Active Ratio but, on the other hand, has the worst results in terms of the neighbour discovery. For example, FPP has the optimal Active Ratio among all the other methods, but it also has the worst EQOS. So it is not a good solution for many types of networks due to it having a high end-to-end delay. So in this paper, the QER is defined that includes the EQOS and the Active Ratio as a unified metric. Based on the QER, the performance of quorum systems can easily be evaluated, noting that larger values of this new metric are desirable. In the next section, the comparisons of the results obtained with these novel methods with the other existing methods based on the QER are presented.

### 5.2.3. The QER

In this section, the results of the proposed methods with the other methods are presented in terms of the QER. As Figure 17 shows, the LPS-Grid($t \times w$) has significantly better QER values than the other methods. It means that the LPS-Grid($t \times w$) makes a better trade-off between the EQOS and the Active Ratio.

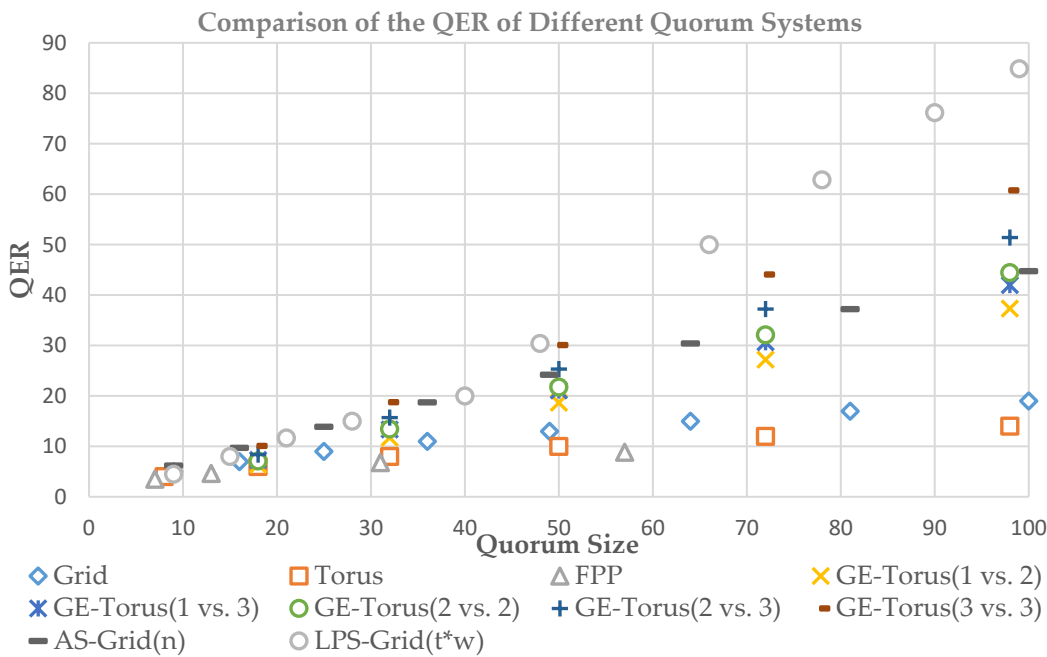

**Figure 17.** The comparison of the QER of the discussed quorum protocols.

### 5.2.4. The Adaptive Mode of the AS-Grid ($t \times w$)

Energy holes emerge in WSNs caused by the feature of data collection when the energy consumption close to a sink is significantly higher than the energy consumption far from the sink. Thereby, the nodes' battery close to the sink drains much faster than other nodes, which causes the death of the network. Utilizing the same quorum system size for different nodes in current quorum-based protocols causes the emergence of energy holes in networks. Therefore, the AS-Grid($t \times w$) uses an adaptive mode whereby nodes can change their quorum system size based on their remaining energy and traffic conditions. In this section, a systematic approach is proposed for the nodes to change their quorum system size dynamically with the varying conditions. To implement the adaptive mode, a parameter $k$ is introduced ($k$ must be an integer number): when nodes decide to change their quorum system size based on their remaining energy or traffic conditions, they can simply add $k$ to $w$ to increase their quorum system size (decrease the Active Ratio) and thus save more energy.

The amount of $k$ that is chosen is based on network conditions and the node's remaining energy adaptively so there is no prior fixed value for $k$ that can be used in advance. In the following, the changes in the value of the EQOS and the Active Ratio of the AS-Grid($t \times w$) is derived as the quorum system size increases by the value of $k$. If it is assumed that the nodes change their quorum system size from $w_1$ to $w_2$ (as assumed before, $t$ is a fixed value for all the nodes in the same network) then:

Increases in the EQOS when $w_2 > w_1$ and $w_2 - w_1 = k$:

$$\frac{t(t + w_2 - 1) + \sum_{j=3}^{t} \sum_{i=0}^{j-3}(t-i) + \sum_{j=2}^{t} \sum_{i=0}^{t-j}(t-i) + \sum_{i=0}^{t-2}(t-i)}{t^2}$$

$$- \frac{t(t + w_1 - 1) + \sum_{j=3}^{t} \sum_{i=0}^{j-3}(t-i) + \sum_{j=2}^{t} \sum_{i=0}^{t-j}(t-i) + \sum_{i=0}^{t-2}(t-i)}{t^2}$$

$$= \frac{t(t + w_2 - 1)}{t^2} - \frac{t(t + w_1 - 1)}{t^2} = \frac{w_2 - w_1}{t} = \frac{k}{t} \qquad (13)$$

Decreases in the Active Ratio when $w_2 > w_1$ and $w_2 - w_1 = k$:

$$\frac{t + w_1 - 1}{t \times w_1} - \frac{t + w_2 - 1}{t \times w_2} = \cdots = \frac{t(w_2 - w_1) + (w_1 - w_2)}{tw_1 w_2} = \frac{tk - k}{tw_1 w_2} = \frac{k(t-1)}{tw_1 w_2} \qquad (14)$$

So with increases in quorum system size ($w$) by the value of $k$, the EQOS is decreased by the value of $\frac{k}{t}$ if $k = 1$ then $\frac{1}{t}$ and the Active Ratio is decreased by the value of $\frac{k(t-1)}{tw_1 w_2}$ if $k = 1$ then $\frac{t-1}{tw_1 w_2}$.

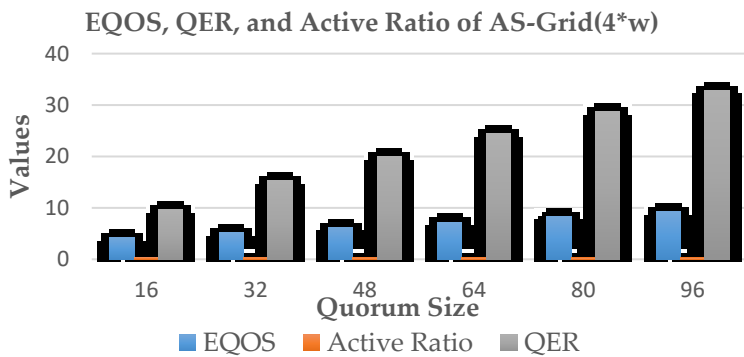

a) $t$=4, $w_1$=4

$EQOS$
$$= \frac{t(t + w - 1) + \sum_{j=3}^{t} \sum_{i=0}^{j-3}(t-i) + \sum_{j=2}^{t} \sum_{i=0}^{t-j}(t-i) + \sum_{i=0}^{t-2}(t-i)}{t^2}$$
$$= 4.25$$
$Active\ Ratio = \frac{t + w - 1}{tw} = 0.4375$

b) $t$=4, $w_2$=5

$EQOS = 4.25 + \frac{w_2 - w_1}{t} = 4.5$
$Active\ Ratio = 0.4375 - \frac{t(w_2 - w_1) + (w_1 - w_2)}{tw_1 w_2} = 0.4$

c) $t$=4, $w_3$=6

$EQOS = 4.5 + \frac{w_3 - w_2}{t} = 4.75$
$Active\ Ratio = 0.4 - \frac{t(w_3 - w_2) + (w_2 - w_3)}{tw_2 w_3} = 0.375$

**Figure 18.** Increase/Decrease in the amount of the EQOS and the Active Ratio in AS-Grid(4×w) with different sizes of $w$.

Figure 18 shows the amount of the EQOS and the Active Ratio in AS-Grid(4×w) with different sizes of $w$. Figure 19 shows the EQOS and the Active Ratio of AS-Grid($4 \times w$) with different sizes of $w$ for system size $n = 0, \dots, 100$. As it is clear, with an increase in the size of an array, the amount of the EQOS is increased and the amount of the Active Ratio is decreased.

**EQOS, QER, and Active Ratio of AS-Grid(4*w)**

**Figure 19.** The EQOS and the Active Ratio of AS-Grid($4 \times w$) with different sizes of $w$ for system size $n = 0, \dots, 100$.

We assume that each node can acquire its remaining energy information from the physical layer. In AS-Grid, we assume that a node adjusts its quorum size according to the following rule, where RE denotes the remaining energy of nodes and $f$ denotes the full energy level of the nodes and $t, w, n, k$ must be an integer:

$$\text{System size} = \begin{cases} t \times w_1, if & f - k < RE < f, \\ t \times (w_1 + 1), if & f - 2k < RE < f - k, \\ & . \\ & . \\ & . \\ t \times (w_1 + nk), if & f - (n+1)k < RE < f - nk \end{cases} \tag{15}$$

The main advantage of the AS-Grid is that there are no fixed thresholds for nodes to change their system size. Therefore, the nodes can change their system size gradually, based on their remaining energy and traffic load. The value of $k$ can be defined by network conditions and the characteristics of the nodes and may differ from one network to another.

## 6. Conclusion

This paper has addressed the active ratio and neighbour discovery in asynchronous ad hoc networks. Two quorum-based power-saving protocols have been identified to maximize the EQOS (neighbour discovery) and minimize the Active Ratio (power consumption). The proposed protocols satisfy the rotation closure property and can be applied to an asynchronous power-saving protocol for MANETs and WSNs. A new metric has also been proposed, the QER, for evaluating quorum-based systems in terms of the Active Ratio and the EQOS. The QER has also been derived for all the discussed quorum systems. An adaptive pattern has been developed for the AS-Grid quorum protocol which can be applied to the clustered wireless sensor networks for each node having a different condition like the remaining energy. The AS-Grid allows each node to choose a different quorum system according to its remaining energy dynamically. Extensive theoretical results have been presented to compare and evaluate all the discussed protocols objectively. Moreover, we believe there is room for improvement, particularly for the end-to-end delay and power consumption vs discovery trade off – these need careful consideration for future neighbour discovery methods.